\documentclass{aastex}          
\usepackage{spr-astr-addons}
\begin{document}
\title{Asymptotic Persistence of Langmuir modes in Kinematically Complex Plasma Flows}
\shorttitle{Persistence of Plasma Waves in the Heliosphere and Beyond}
\shortauthors{Arabuli \& Rogava  }
\author{Ketevan Arabuli} 
\affil{KU Leuven, Leuven, Belgium}
\author{Andria Rogava} 
\affil{Institute for Theoretical Physics, Ilia State University, G. Tsereteli 3, 0162 Tbilisi, Georgia}


\begin{abstract}
Dynamics of Langmuir modes - Langmuir waves (LW) and shear Langmuir vortices (SLV) - in kinematically complex astrophysical plasma flows is studied. It is found that they exhibit a number of peculiar, velocity shear induced, {\it asymptotically persistent} phenomena: efficient energy exchange with the background flow, various kinds of instabilities, leading to their exponential growth; echoing solutions with persistent wave-vortex-wave conversions. Remarkable similarity of these phenomena with ones happening with compressible acoustic modes, revealed in \citep{Mahajan_Rogava_1999} is pointed out. The relevance and possible importance of these phenomena for different types of astrophysical plasma flow patterns with kinematic complexity is discussed. In particular, we argue that these physical processes may account for the persistent appearance of plasma oscillations in the heliosphere and in interstellar plasma flows. In particular, we believe that kinematically complex motion of plasma may naturally lead to asymptotically persistent appearance of Langmuir modes born, grown, fed, sustained and maintained by these flows.
\end{abstract}




\section{Introduction}

It is widely known that both the wave and the vortical modes of motion are common in many different kinds of astronomical objects. Since the visible universe is predominantly made of plasma it is obvious that the study of both the wave and the vortical motions in plasma flows is of key importance in theoretical astrophysics. In a larger perspective, for several decades, vortex dynamics is widely recognized as a fundamental iterdisciplinary scientific problem \citep{Saffman_1993}. Remarkably, there is a considerable degree of similarity between the vortices in plasmas and neutral fluids \citep{Petviashvili_1980, Horton_Hasegawa_1994}. Usually vortical motions in plasmas are associated with complicated, highly nonlinear systems. But with the gradual disclosure of so called {\it nonmodal} phenomena in neutral fluid shear flows \citep{Kelvin_1887, Goldreich_Lynden-Bell_1965, Trefethen_1993} it became increasingly clear that even on the linear level there exists a number of of interesting shear-induced effects associated with  the vortex and wave linear dynamics both in neutral fluid and plasma shear flows \citep{Mahajan_Rogava_1999}. These phenomena are characterized by intense energy exchange between the wave and the vortical modes and the background flow.

The importance of velocity shear induced phenomena in astrophysics is widely known, acknowledged and extensively discussed for several decades in a broad variety of important classes of astrophysical flows. The incomplete list of astronomical situations, where these phenomena have been considered includes: density waves in galaxies \citep{Goldreich_Lynden-Bell_1965, Toomre_1969, Goldreich_Tremaine_1978, Fan_Lou_1997, Rogava_1999, Poedts_Rogava_2002}; transiently growing incompressible modes in accretion discs \citep{Lominadze_1998, Balbus_Hawley_1992, Tagger_1992}; reciprocal transformations of solar magnetohydrodynamic (MHD) waves and their role in the acceleration of the solar wind \citep{Poedts_1998, Rogava_2000}. It was also found out that within relativistic pulsar magnetospheric $e^+ e^-$ plasmas transformation of non-escaping longitudinal Langmuir modes into escaping electromagnetic waves \citep{Mahajan_1997} may happen and contribute to the generation of pulsar radio emission. Another large class of astrophysical flows with highly nontrivial velocity inhomogeneity are the galactic and extragalactic jets \citep{Lovelace_1987, Ferrari_1998, Chen_Zhang_2021}. Their kinematic complexity is caused and ensured by complex spatial geometry of ordered magnetic fields, responsible for the jet collimation and efficient acceleration of matter up to relativistic velocities \citep{Lovelace_1991, Rogava_Khujadze_1997}. 

As a rule spatial inhomogeneity of astrophysical flows is quite complicated, with different components of the velocity field possessing different, often nonlinear gradients. Sometimes, when considered physical processes  have length scales $l$ that  are sufficiently small compared with the outward spatial length-scales $L$, the flows may be assumed to have locally linear field-profile geometries. It is well-known that the $l \ll L$ approximation is elemental for the efficacy of commonly used Kelvin's method \citep{Kelvin_1887} of change of variables. However the validity of the approximation is often questionable in astrophysical situations. In many cases of astrophysical importance shear flows hardly ever are that simple. 

The dazzling variety of flow patterns in the solar atmosphere and the heliosphere, for example, is characterized by a notable degree of {\it kinematic complexity}. \citep{Mahajan_Rogava_1999}. Convective motions within the solar photosphere, for instance, are characterized by highly nontrivial, non-potential velocity fields coupled with the additional effect of the solar differential rotation. Further up, according to \citet{Parker_1974}, magnetic filaments being rooted in the convective zone, with their footpoints continuously squeezed by the convective granules, leads to kinematically complex, variable plasma motion within these flux tubes \citep{Rogava_2000}. There is also an abundant and ever-increasing observational evidence \citep{pneuman_Orral_1980, Pike_Mason_1998} for the presence of significantly complex transient and/or oscillatory motions within coronal plasma loops, spicules and solar tornadoes. These motions are superimposed with the motions within horizontal fibrils and with downflowing streams, constituting "complex flow networks" \citep{ath86} in the transition region. It was repeatedly argued \citep{Poedts_1998, Mahajan_Rogava_1999, Rogava_2000} that presence of these kinematically complex patterns of plasma flows, coupled with the presence of magnetic fields, often governing these motions, should heavily influence modes of collective behavior, generation and propagation of waves and vortices in the solar atmosphere, possibly contributing to the coronal heating, and to the acceleration of the solar wind. 

Bearing in mind the necessity and importance of nonmodal studies of modes of collective behavior in astrophysical shear flows with multidimensional geometry and nontrivial kinematics \citep{Mahajan_Rogava_1999} developed a non-asymptotic method apt to inspect the response of such flows to small-scale perturbations. It allows to reduce the initial value problem to a set of manageable ordinary differential equations in time. A similar method, designed for flows with spatially uniform shearing rates, had been used in hydrodynamics \citep{Lagnado_1984, craik_criminale_1986}. These studies showed a remarkable richness in the  spectrum of both incompressible and compressible fluctuations sustained by these flow. The most impressive feature of revealed exotic phenomena was their {\it asymptotic persistence} in the absence of viscosity. The study revealed the presence of ``Echoing", strongly  unstable solutions, efficient energy exchange between the ``parent" flow and the waves and instabilities sustained by it, mode conversions and beat wave phenomena \citep{Mahajan_Rogava_1999}. Further application of this method to kinematically complex electrostatic \citep{Osmanov_2015} and magnetohydrodynamic \citep{Osmanov_2012} flows showed that these phenomena could be of relevance in cosmic plasma flows as well. 

The presence of the velocity shear brings to existence new modes of collective behavior; Kelvin modes \citep{Kelvin_1887} are the earliest example discovered more than a century ago. Considering linear evolution of electrostatic perturbations in a cold, unmagnetized, two-component plane-parallel plasma shear flow it was found out \citep{Rogava_1998} that similar modes may exist in electrostatic plasma flows. It was shown that the velocity shear induces, due to the non-normality of the linear dynamics, a new elementary mode of plasma {\it non-periodic} collective behavior the SLV, associated with nonoscillatory motion of plasma species. The study showed that SLV may efficiently exchange energy  with the flow. It turned out that the two-dimensional SLV could extract the flow energy only transiently, while the three-dimensional ones could exhibit asymptotic persistence. It was also discovered that in moderate and high velocity inhomogeneity (velocity shear) regimes, the remarkable non-resonant conversion of SLV into LW could take place.

The aim of this paper is to study dynamics of LW and SLV in the context of astrophysical shear flows with kinematic complexity. Recent observational developments with the study of plasma waves and instabilities within and beyond the heliosphere strengthens our confidence that theoretical studies of Langmuir modes in complex flows may find interesting applications in the astronomical context. These groundbreaking observations showed that even beyond the limits of the solar system, beyond the heliosphere and outside the heliopause boundary, LW are present. In particular, Voyager 1, the first ever in situ probe of the interstellar medium, very recently discovered the distinct presence of LW; it was reported \citep{Ocket_2021} that from 2017 onwards the probe's Plasma Wave System detected a class of weak but persistent, narrowband plasma wave emission in its data. Most remarkably large-scale density fluctuations were discovered, tracing possible presence of interstellar turbulence between episodes of previously detected plasma oscillations. The persistence of this emission indicated that the Voyager 1 was able to detect the interstellar plasma density variations even in the absence of shock-generated LW events.
  
The paper is arranged in the following manner: In the Main Consideration, based on results of \citep{Rogava_1998} and \citep{Mahajan_Rogava_1999}, we develop the linear theory of Langmuir modes in kinematically complex plasma flows and derive an explicit second-order ODE governing the evolution of these modes. In the Discussion we study the evolution of these modes and find out that there is a remarkable similarity between the evolution of Langmuir modes and acoustic modes \citep{Mahajan_Rogava_1999}. Our study makes clear that Langmuir modes in kinematically complex flows exhibit an impressive variety of peculiar, shear-induced, {\it asymptotically persistent} phenomena, involving (but not limited to) various kinds of, including parametrically driven, instabilities,  efficient energy exchange processes with the background flow, interesting ``echoing" dynamics of perturbations, wave-vortex-wave conversion events and appearance of beat wave phenomena. In the concluding section of the paper we discuss the possible relevance and connection of the "asymptotic persistence" with the surprisingly persistent appearance of Langmuir modes within the heliosphere.



\section{Main Consideration}

In this paper we employ the general approach developed in \citep{Mahajan_Rogava_1999}: we consider only small-scale perturbations with their length-scales $l_i$ much smaller than the characteristic scales $L_i$ of the flow ($i=x,y,z$). When $l_i{\ll}L_i$ the spatial inhomogeneity of an arbitrary background velocity field ${\bf U}({\bf r})$ in the close neighbourhood of a point $A$ with a radius vector ${\bf r}_A$  ($|{\bf r}-{\bf r}_A)|/|{{\bf r}_A}| {\ll}1$) may be approximated by the linear terms in its Taylor expansion: $U_i(x_i)\simeq {U_A}_i+ a_{ik}x_k$. It leads to the definition of the {\it Shear Matrix} \citep{Mahajan_Rogava_1999}:
\begin{equation}
{\cal S} \equiv (a_{ik}) = 
\begin{pmatrix}
U_{x,x} & U_{x,y} & U_{x,z} \\
U_{y,x} & U_{y,y} & U_{y,z} \\
U_{z,x} & U_{z,y} & U_{z,z} 
\end{pmatrix}
\end{equation}

For flows with uniform equilibrium density, the velocity field is {\it solenoidal} ($\nabla\cdot{\bf U}=0$), implying that the Shear Matrix has to be traceless. The linearized convective derivative is reduced to ${\cal D}u_i+a_{ik}u_k$, where ${\cal D}\equiv\partial_t+U_i(x,y,z) \partial_i$. The essential idea of this approach is to find a proper form of an unknown function that would facilitate solution, viz. that would {\it eliminate} the spatial dependence in the operator ${\cal D}$. It turns out that for the {\it ansatz}:
\begin{equation}
F({\bf r},t)\equiv\hat F({\bf k}(t),t) e^{i\varphi}, 
\end{equation}
\begin{equation}
{\varphi}({\bf r},t)\equiv {\bf k} \cdot {\bf r} -{\bf U_A} \cdot {\int}_0^{t}{\bf k}(t')dt',
\end{equation}
the elimination does happen, the {\it ansatz} ensures that the linearized convective derivative reduces to an ordinary derivative in time for the amplitude of the perturbation: 
\begin{equation} 
{\cal D}F=e^{i\varphi} \partial_t \hat F^,
\end{equation}
but it occurs if and only if the wave vector ${\bf k}$ acquires the time-dependence given by:
\begin{equation}
\partial_t {\bf k}+{\cal S}^T\cdot{\bf k}=0, 
\end{equation}
where ${\cal S}^T$ stands for the transposed shear matrix \citep{craik_criminale_1986, Mahajan_Rogava_1999}. 

Consequently the set of partial differential equations (PDE's), describing the evolution of linear perturbations reduces to the set of non-autonomous ordinary differential equations (ODE's). It dramatically simplifies the task, reducing mathematically challenging sets of PDE's (often with complicated and obscure boundary conditions) to the initial value problem formulated for the set of non-autonomous ODE's. It was repeatedly shown \citep{Mahajan_Rogava_1999, Rogava_2001, Rogava_2003a, Rogava_2003b, Rogava_2007, Osmanov_2012, Osmanov_2015} that the latter can be inspected and reveal previously overlooked modes of collective behaviour, sustained and modified by the velocity fields with kinematic complexity.

Following \citep{Rogava_1998} we consider nonrelativistic, quasineutral, cold and unmagnetized fluid plasma. That is, our main assumptions are standard for the study of plasma oscillations. We do assume there is no magnetic field  $(\bf B = 0)$; there are no thermal motions $(T=0)$; we adopt ``no boundaries'' condition: the plasma is infinite in extent; and the oscillations are electrostatic: $\vec E = - \nabla \varphi$. 

For the mean flow velocity we adopt the two-dimensional kinematically complex flow model used in \citep{Mahajan_Rogava_1999}: ${\bf U}(x,y)\equiv U_x (x,y){\bf e}_x+U_y(x,y){\bf e}_y$ with $a_{11}=-a_{22}\equiv\sigma$,
$a_{12}\equiv a$ and $a_{21}\equiv b$. Applying the ansatz (2-5) we derive the set of linearized, first order ODE's:  
\begin{equation} 
\varrho^{(1)}={\cal K}_x v_x+{\cal K}_y v_y,
\end{equation}
\begin{equation}
v_x^{(1)} = -\varepsilon v_x - R_1v_y - (W/ {\cal K})^2 {\cal K}_x\varrho,
\end{equation}
\begin{equation}
v_y^{(1)} = - R_2v_x + \varepsilon v_y -(W/ {\cal K})^2  {\cal K}_{y} \varrho.
\end{equation} 
while the dimensionless components of the wavevector ${\cal K}_{x,y}\equiv k_{x,y}/k_x(0)$ vary in time as:
\begin{equation}
{\cal K}_x^{(1)} = - \varepsilon{\cal K}_x - R_2{\cal K}_{y},
\end{equation}
\begin{equation}
{\cal K}_{y}^{(1)} = - R_1{\cal K}_x + \varepsilon{\cal K}_{y}. 
\end{equation}

In these equations we introduced the dimensionless notation: $W \equiv \omega / c k_x(0)$, ($\omega \equiv (4 \pi q^2 N/m)^{1/2}$ is the plasma frequency); $\varepsilon \equiv \sigma/ck_x(0)$, $R_1 \equiv a/ck_x(0)$, $R_2 \equiv b/c k_x(0)$, , $\varrho \equiv i({\hat \rho}'/\rho_0)$, and $v_{x,y} \equiv \hat u_{x,y}/c$. Note that $f^{(n)}$ denotes n-th order time derivative of $f$ by the dimensionless time variable $\tau \equiv ck_x(0)t$.

These equations, after taking one more time derivative, readily lead to:
\begin{equation}
 \begin{Bmatrix}
{\cal K}_x \\
{\cal K}_y 
\end{Bmatrix}^{(2)} = {\Lambda}^2
 \begin{Bmatrix}
{\cal K}_x \\
{\cal K}_y 
\end{Bmatrix},
 \end{equation}
where ${\Lambda}^2 \equiv {\varepsilon}^2 + R_1 R_2$. 

Another interesting and important property of the dimensionless wavevector is that the z-component of the vector ${\bf k}^{(1)} \times {\bf k}$ is a conserved quantity:
\begin{equation}
\Delta  = {\cal K}_x^{(1)} {\cal K}_y - {\cal K}_y^{(1)} {\cal K}_x = const.
 \end{equation}

Introducing in the k-space the polar angle $\varphi(t)$, implying ${\cal K}^2 \equiv {\cal K}_x^2 + {\cal K}_x^2$, ${\cal K}_x = {\cal K} cos {\varphi}$, and {\cal K}$_y = {\cal K} sin {\varphi}$, we can easily find out that:
\begin{equation}
\Delta  = - {\cal K}^2 \varphi^{(1)} \equiv - {\cal K}^2 \omega = const,
 \end{equation}
where, obviously, $\omega = \varphi^{(1)}$ is the angular velocity of the wavevector rotation in the k-space! 

Evidently ${\cal K}$ plays an important role in the mathematical description of the Langmuir mode dynamics. Therefore, it is important to study in detail its properties. We can easily show that:  
\begin{equation}
({\cal K}^2)^{(1)} = 2  \left[\varepsilon ( {\cal K}_y^2 -{\cal K}_x^2) - (R_1 + R_2) {\cal K}_x {\cal K}_y \right].    
\end{equation}

Furthermore, with some additional calculations and taking into account (11) and (12) we derive:
\begin{equation}
({\cal K}^2)^{(2)} = 4\Lambda^2 {\cal K}^2 + 2 \Delta (R_1 - R_2).    
\end{equation}
\begin{equation}
{\cal K}^{(1)} = \sqrt{2 \Lambda^2 {\cal K}^2 + \Delta(R_1 - R_2) - {\cal K}  {\cal K}^{(2)}},
 \end{equation}
\begin{equation}
\frac{{\cal K}^{(2)}}{{\cal K}} = \Lambda^2 + \frac{\Delta^2}{{\cal K}^4},
 \end{equation}
 while combining the last two equations we finally derive an explicit first order ODE for the ${\cal K}^2$ function:
 \begin{equation}
 ({\cal K}^2)^{(1)} = \sqrt{ {\Lambda^2 {\cal K}^4 + \Delta (R_1 - R_2){\cal K}^2 - \Delta^2} }.
 \end{equation}

 This is important equation because it allows to determine the function $X(t) \equiv {\cal K}^2$ by calculating analytically the well-known integral: 
\begin{equation}
\int \frac {dX} {\sqrt{A X^2 + B X + C}} = 2 \int dt,  
\end{equation}
where $A \equiv \Lambda^2$, $B \equiv (R_1 - R_2)\Delta$, and $C \equiv - \Delta^2$. Depending on values of these parameters the function $\cal K$ shall vary in time in a polynomial, periodic or exponential way. The similar result, of course, is achieved if one solves for ${\cal K}_x$ and ${\cal K}_y$ and determines the variability of the $\cal K$ via its definition.

Turning our attention now to the set (6-8) we can easily prove that the following combination of variables is a conserved quantity:
\begin{equation}
{\cal C} \equiv {\cal K}_y v_x - {\cal K}_x v_y + (R_1 - R_2) \varrho  = const  
\end{equation}

Subsequently, calculating ${\varrho}^{(2)}$. We find out that: 
\begin{equation}
{\varrho}^{(2)} + W^2 \varrho = 2 ({\cal K}_x^{(1)} v_x + {\cal K}_y^{(1)} v_y) 
\end{equation}

It is possible to calculate the velocity perturbation components in terms of the density perturbation and its first derivative. It is straightforward to derive that:
\begin{equation}
v_x = \left( {\cal K}_y[{\cal C} + (R_2 - R_1) {\varrho}] + {\cal K}_x {\varrho}^{(1)} \right)/{\cal K}^2, 
\end{equation}
\begin{equation}
v_y = \left( - {\cal K}_x[{\cal C} + (R_2 - R_1) {\varrho}] + {\cal K}_y {\varrho}^{(1)} \right)/{\cal K}^2. 
\end{equation}

Obviously (21-23) can be used to derive explicit, second-order nonautonomous ODE for the density perturbation. Taking into account the properties of the $\cal K$ function we come to the result: 
\begin{equation}
\varrho^{(2)} - {{({\cal K}^2)^{(1)}}\over{{\cal K}^2}} \varrho^{(1)} + \left[W^2 + 
{{2 \Delta (R_1-R_2)}\over{{\cal K}^2}}\right] \varrho = {{2  {\cal C} \Delta}\over{{\cal K}^2}}
\end{equation}

By eliminating the first derivative term $\varrho^{(1)}$ this equation can be rewritten in a more elegant and concise way. The task is fulfilled by introducing yet another auxiliary variable $\Psi \equiv \varrho/{\cal K}$. It is straightforward to see that for this function we will have the following equation:
\begin{equation}
{\Psi}^{(2)} + {\Omega}^2 {\Psi} = {{2 {\cal C} {\Delta} }\over{{{\cal K}^3}}}
\end{equation}
where 
\begin{equation}
{\Omega}^2 \equiv W^2 + {{{\cal K}^{(2)}}\over{{\cal K}}} - {{2 ({\cal K}^{(1)})^2}\over{{\cal K}^2}} + {{2 \Delta (R_1 - R_2)}\over{{\cal K}^2}}.
\end{equation}

\begin{figure}
	\includegraphics[width=\columnwidth]{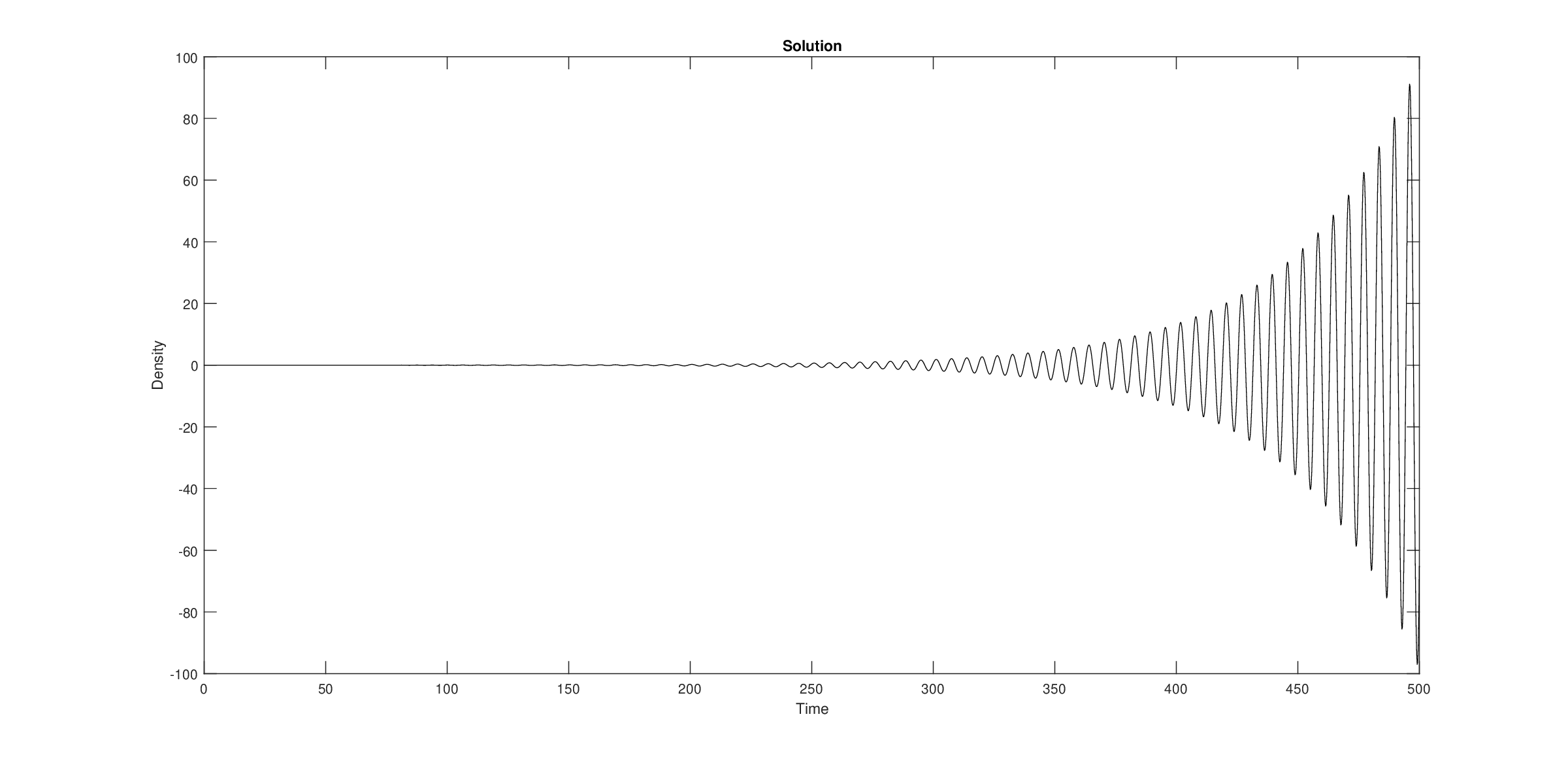}
    \caption{Exponentially growing density perturbation $\varrho$, for $R_1 = 2 \cdot 10^{-3}$, $R_2 = 0.2$, $\varepsilon = 0$, $K_y(0) = 0.1$, $C = 0$, $W = 1$, ${\Psi}(0) = 10^{-2}$, and ${{\Psi}^{'}}(0) = 0$.}
    \label{fig:example_figure1}
\end{figure}
\begin{figure}
	\includegraphics[width=\columnwidth]{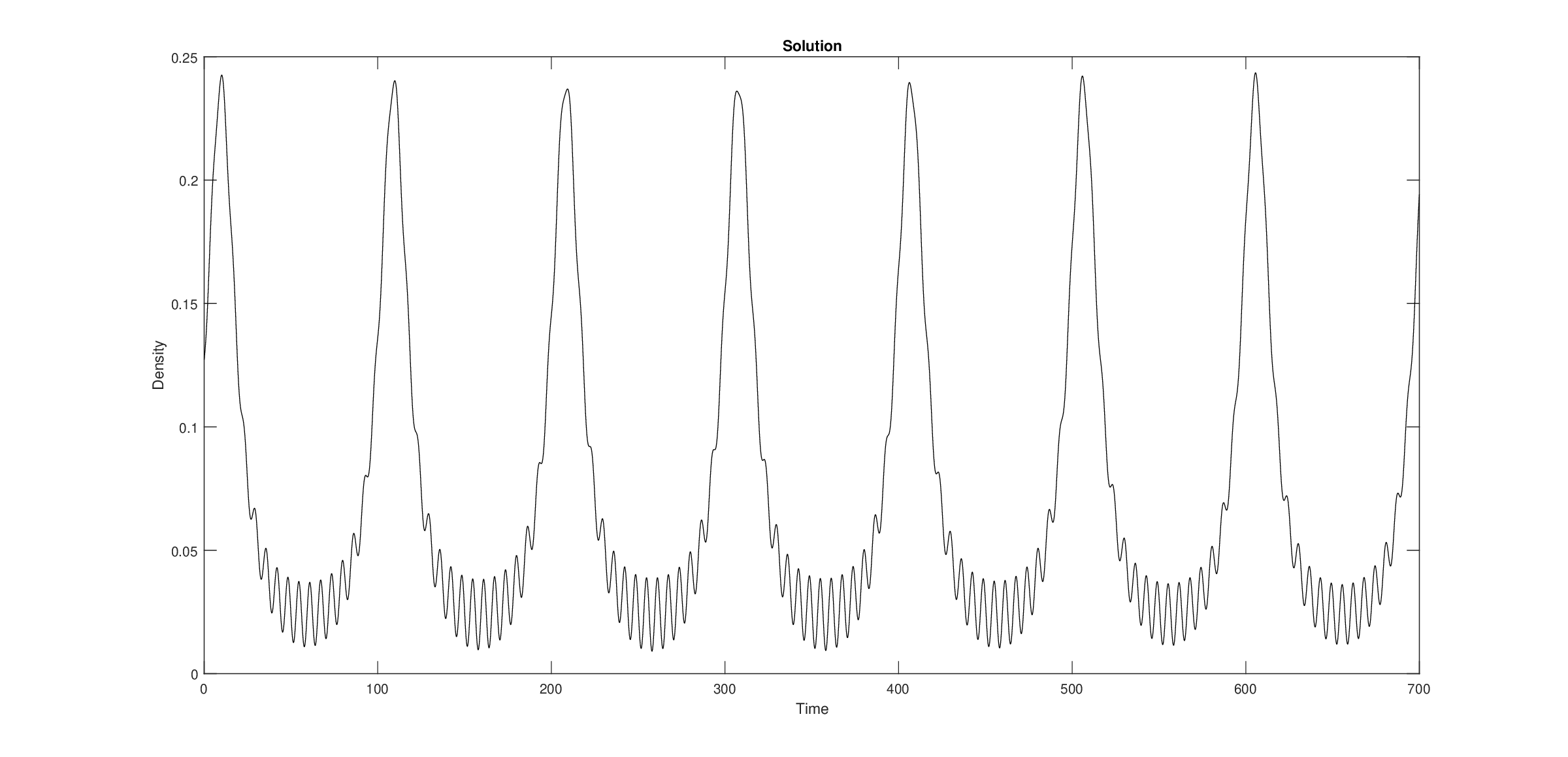}
    \caption{Asymptotic persistence of echoing SLV interacting with LW for the density $\varrho$ perturbations. $\varepsilon = 0$, $R_1 = 0.1$, $R_2 = -0.01$, ${\cal C} = 1.2 $, $W = 1$, ${\cal K}_y(0) = 1$, $\Psi(0) =9 \cdot 10^{-2}$, and ${{\Psi}^{'}}(0) = 8.66 \cdot 10^{-3}$. }
    \label{fig:example_figure2}
\end{figure}

Which can be further simplified by taking into account (16) and (17). The final result is: 
\begin{equation}
{\Psi}^{(2)} + {\left[ W^2 - \Lambda^2 + \frac{3\Delta^2}{{\cal K}^4} \right]}\Psi = \frac{2 {\cal C}  \Delta}{{\cal K}^3}
\end{equation}

This equation is very convenient for inspecting and classifying possible regimes of Langmuir mode dynamics. It is important to note that it features a striking similarity with the equation (7) from \citep{Mahajan_Rogava_1999}:
\begin{equation}
{\Psi}^{(2)} + {\left[ {\cal K}^2 - \Lambda^2 + \frac{3\Delta^2}{{\cal K}^4} \right]}\Psi = \frac{2 {\cal C}  \Delta}{{\cal K}^3},
\end{equation}
where dynamics of acoustic modes in kinematically complex flows was considered. The only, but quite important, difference between these two equations is related with the "generic" frequency of oscillations: for acoustic perturbations it is equal to ${\cal K}^2$, while for Langmuir modes, obviously, instead we do have the square of the normalized plasma oscillation (Langmuir) frequency $W^2$.

The analogy suggests classes of probable evolution scenarios for the evolution of Langmuir modes in kinematically complex flows. It is evident that, similarly to the acoustic case, the value and the magnitude of the $\Lambda^2$ parameter would determine temporal evolution of the $\cal K$ function, determining, in its turn, the time-dependence of the effective frequency $\Omega$ and the inhomogeneity term on the right hand side of the (27). When $\Lambda^2=0$ the evolution is fully identical to the case of plane-parallel flows studied in detail in \citep{Rogava_1998}. When $\Lambda^2 > 0$, the time-dependence of the $\cal K$ function is exponential, transcribing into shear-induced, exponentially growing waves within kinematically complex flows \citep{Mahajan_Rogava_1999}. We have found analogous, exponentially growing LW in our case, represented by the \ref{fig:example_figure1}. The values of parameters here are: $R_1 = 2 \cdot 10^{-3}$, $R_2 = 0.2$, $\varepsilon = 0$, $K_y(0) = 0.1$, $C = 0$, $W = 1$, $\Psi(0) = 10^{-2}$, and ${{\Psi}^{'}}(0) = 0$. Of course viscous damping would eventually "kill" these perturbations on a nonlinear stage of their evolution, but the existence of robust, shear-induced linear exponential growth indicates at the important ability of LW to be asymptotically persistent in flows with kinematic complexity. When $\Lambda^2 < 0$ we found that nonperiodic modes of plasma collective behaviour - shear Langmuir vortices (SLV) - exhibit remarkable ``echoing'' behaviour, reminiscent again of the analogous behaviour of Kelvin modes in hydrodynamic flows \citep{Mahajan_Rogava_1999}. It is featured on \ref{fig:example_figure2}. The values of parameters here are: $\varepsilon = 0$, $R_1 = 0.1$, $R_2 = -0.01$, ${\cal C} = 1.2 $, $W = 1$, ${\cal K}_y(0) = 1$, $\Psi(0) =9 \cdot 10^{-2}$, and ${{\Psi}^{'}}(0) = 8.66 \cdot 10^{-3}$. And again and again, this is clear example of asymptotic persistence - SLV and LW keep converting into one another and it persists for a long intervals of time. In the $\Lambda^2 < 0$ the existence of parametrically unstable LW is also expected, resembling similar instability disclosed for acoustic waves in \citep{Mahajan_Rogava_1999}. It is one of topics of our future research to determine those isolated areas of the k-space, where parametrically unstable LS are expected to exist.

Evidently the problem is quite complex, the equations contain a number of free parameters related to the physics of the Langmuir modes and the kinematic portrait of the background flow. The thorough and systematic exploration of all interesting regimes, with certain cases allowing exactly or approximately analytic solutions, is currently undergoing and the results will be published in a forthcoming, separate publication. 


\section{Discussion and Conclusions}

Within last few decades studies of the phenomenon of {\it asymptotic persistence} \citep{Mahajan_Rogava_1999} of shear-induced mode dynamics in kinematically complex flows amply demonstrated that these processes take place in a wide variety of physical or astrophysical systems. At the other hand, even in plane-parallel flows Langmuir modes are known to exhibit impressive variety of nonmodal phenomena \citep{Rogava_1998}, including exponential growth of LW and onset of  totally new, shear-flow-generated nonperiodic modes of collective behaviour - {\it Shear Langmuir Vortexes (SLV)}. It was found out that LW and SLV efficiently interact with background flows and may convert into one another. The main purpose of the current study was to fuse these two issues, to make one step further and check what happens with Langmuir modes (LW and SLV) in plasma flows with kinematic complexity. Naturally the main field of application that we have for starters in mind is solar physics, namely highly probable relevance of this physical study to the dynamics of plasma modes in the heliosphere and beyond the heliopause, in the interstellar medium. 

The relevance and importance of this context seems very well-established. Empirical study of LW in the heliosphere has undergone significant advancements in recent years. Recently it became increasingly clear that these waves, intrinsically connected to electron dynamics, may play an important role in providing a channel of the energy transfer from the Sun to the outermost regions of the heliosphere \citep{Briand_2015}. Besides, the measurable growth rates and saturation levels of LW, processes of fundamental nature, potentially provide credible diagnostic capabilities. In addition, in the heliosphere, an almost collisionless medium, LW may play the role of collisions, participating in the exchange of energy between different species. 

On the basis of this  understanding recent empirical research has expanded the study objectives of LW to a number of contexts, in both quiescent conditions of the solar wind and during transient events, including planetary environments such as the ionosphere, the magnetotail, and the foreshock, interaction of the heliospheric plasma with ion and dust tails of rapidly moving comets, as well as the processes well beyond the heliopause, in the interplanetary medium. Namely, 
several spacecrafts, including IMP-6, IMP-7, IMP-8, Helios, STEREO, and Wind, have observed electron events, associated with LW and type III radio bursts as far out as 1 AU. These observations have been critical in understanding the characteristics and dynamics of LW \citep{Lorfing_2023}.

Furthermore, Solar electron beams responsible for type III radio emission generate LW as they propagate out from the Sun. They are observed via in-situ electric field measurements. These LW are not smoothly distributed but appear in discrete clumps, commonly attributed to the turbulent nature of the solar wind electron density \citep{Reid_Kontar_2017}. Still there is no exact knowledge of the interaction between the density turbulence and the LW electric fields, the issue is only qualitatively understood. Namely, on the basis of the weak turbulence simulations, it was checked out how solar wind density turbulence might changes the probability distribution functions, mean value, and variance of the beam-driven electric field distributions.

Concluding the paper we can only repeat that our main purpose in this study was to see whether asymptotic persistence of Langmuir modes in kinematically complex shear plasma flows could be related to the observed persistency of plasma waves in the heliosphere and beyond, in the interstellar space immediately beyond the heliopause. Our first results seem very encouraging and pointing out at the highly probable relevance of the assumption. Needless to say further study is required to prove the validity of our conjecture in a more comprehensive way. 

\acknowledgments
Our research was supported in part by Shota Rustaveli National Science Foundation of Georgia (SRNSFG) grant FR-18-14747. Andria Rogava wishes to thank for hospitality Centre for mathematical Plasma-Astrophysics, KU Leuven, where a part of this work was completed.

\bibliographystyle{spr-mp-nameyear-cnd}
\bibliography{main}

\begin{thebibliography}{40}
\ifx \bisbn   \undefined \def \bisbn  #1{ISBN #1}\fi
\ifx \binits  \undefined \def \binits#1{#1}\fi
\ifx \bauthor  \undefined \def \bauthor#1{#1}\fi
\ifx \batitle  \undefined \def \batitle#1{#1}\fi
\ifx \bjtitle  \undefined \def \bjtitle#1{#1}\fi
\ifx \bvolume  \undefined \def \bvolume#1{\textbf{#1}}\fi
\ifx \byear  \undefined \def \byear#1{#1}\fi
\ifx \bissue  \undefined \def \bissue#1{#1}\fi
\ifx \bfpage  \undefined \def \bfpage#1{#1}\fi
\ifx \blpage  \undefined \def \blpage #1{#1}\fi
\ifx \burl  \undefined \def \burl#1{\textsf{#1}}\fi
\ifx \doiurl  \undefined \def \doiurl#1{\textsf{#1}}\fi
\ifx \betal  \undefined \def \betal{\textit{et al.}}\fi
\ifx \binstitute  \undefined \def \binstitute#1{#1}\fi
\ifx \binstitutionaled  \undefined \def \binstitutionaled#1{#1}\fi
\ifx \bctitle  \undefined \def \bctitle#1{#1}\fi
\ifx \beditor  \undefined \def \beditor#1{#1}\fi
\ifx \bpublisher  \undefined \def \bpublisher#1{#1}\fi
\ifx \bbtitle  \undefined \def \bbtitle#1{#1}\fi
\ifx \bedition  \undefined \def \bedition#1{#1}\fi
\ifx \bseriesno  \undefined \def \bseriesno#1{#1}\fi
\ifx \blocation  \undefined \def \blocation#1{#1}\fi
\ifx \bsertitle  \undefined \def \bsertitle#1{#1}\fi
\ifx \bsnm \undefined \def \bsnm#1{#1}\fi
\ifx \bsuffix \undefined \def \bsuffix#1{#1}\fi
\ifx \bparticle \undefined \def \bparticle#1{#1}\fi
\ifx \barticle \undefined \def \barticle#1{#1}\fi
\ifx \bconfdate \undefined \def \bconfdate #1{#1} \fi
\ifx \botherref \undefined \def \botherref #1{#1} \fi
\ifx \url \undefined \def \url#1{\textsf{#1}} \fi
\ifx \bchapter \undefined \def \bchapter#1{#1} \fi
\ifx \bbook \undefined \def \bbook#1{#1} \fi
\ifx \bcomment \undefined \def \bcomment#1{#1} \fi
\ifx \oauthor \undefined \def \oauthor#1{#1} \fi
\ifx \citeauthoryear \undefined \def \citeauthoryear#1{#1} \fi
\ifx \endbibitem  \undefined \def \endbibitem {}\fi
\ifx \bconflocation  \undefined \def \bconflocation#1{#1} \fi
\ifx \arxivurl  \undefined \def \arxivurl#1{\textsf{#1}} \fi
\csname PreBibitemsHook\endcsname

\bibitem[\protect\citeauthoryear{Athay}{1986}]{ath86}
\begin{bchapter}
\bauthor{\bsnm{Athay}, \binits{R.G.}}:
In: \beditor{\bsnm{Sturrock}, \binits{P.A.}} (ed.)
\bbtitle{Physics of the Sun}
vol. \bseriesno{2},
p. \bfpage{51}
(\byear{1986})
\end{bchapter}
\endbibitem

\bibitem[\protect\citeauthoryear{Balbus and Hawley}{1992}]{Balbus_Hawley_1992}
\begin{barticle}
\bauthor{\bsnm{Balbus}, \binits{S.A.}},
\bauthor{\bsnm{Hawley}, \binits{J.F.}}:
\bjtitle{The Astrophysical Journal}
\bvolume{400},
\bfpage{610}
(\byear{1992})
\end{barticle}
\endbibitem

\bibitem[\protect\citeauthoryear{Briand}{2015}]{Briand_2015}
\begin{botherref}
\oauthor{\bsnm{Briand}, \binits{C.}}:
Journal of Plasma Physics
\textbf{81}(2)
(2015)
\end{botherref}
\endbibitem

\bibitem[\protect\citeauthoryear{Chen}{2021}]{Chen_Zhang_2021}
\begin{barticle}
\bauthor{\bsnm{Chen}, \binits{B.} \bsuffix{L.and~Zhang}}:
\bjtitle{The Astrophysical Journal}
\bvolume{906}(\bissue{2}),
\bfpage{105}
(\byear{2021})
\end{barticle}
\endbibitem

\bibitem[\protect\citeauthoryear{Craik and Criminale}{1986}]{craik_criminale_1986}
\begin{barticle}
\bauthor{\bsnm{Craik}, \binits{A.D.D.}},
\bauthor{\bsnm{Criminale}, \binits{W.O.}}:
\bjtitle{Proc. R. Soc. Lond. A}
\bvolume{406},
\bfpage{13}
(\byear{1986})
\end{barticle}
\endbibitem

\bibitem[\protect\citeauthoryear{Fan and qing Lou}{1997}]{Fan_Lou_1997}
\begin{barticle}
\bauthor{\bsnm{Fan}, \binits{Z.}},
\bauthor{\bsnm{Lou}, \binits{Y.-q.}}:
\bjtitle{Monthly Notices of the Royal Astronomical Society}
\bvolume{291},
\bfpage{91}
(\byear{1997})
\end{barticle}
\endbibitem

\bibitem[\protect\citeauthoryear{Ferrari}{1998}]{Ferrari_1998}
\begin{barticle}
\bauthor{\bsnm{Ferrari}, \binits{A.}}:
\bjtitle{Annual Review of Astronomy and Astrophysics}
\bvolume{36}(\bissue{1}),
\bfpage{539}
(\byear{1998})
\end{barticle}
\endbibitem

\bibitem[\protect\citeauthoryear{Goldreich and Lynden-Bell}{1965}]{Goldreich_Lynden-Bell_1965}
\begin{barticle}
\bauthor{\bsnm{Goldreich}, \binits{P.}},
\bauthor{\bsnm{Lynden-Bell}, \binits{D.}}:
\bjtitle{Monthly Notices of the Royal Astronomical Society}
\bvolume{130},
\bfpage{125}
(\byear{1965})
\end{barticle}
\endbibitem

\bibitem[\protect\citeauthoryear{Goldreich and Tremaine}{1978}]{Goldreich_Tremaine_1978}
\begin{barticle}
\bauthor{\bsnm{Goldreich}, \binits{P.}},
\bauthor{\bsnm{Tremaine}, \binits{S.}}:
\bjtitle{Astrophysical Journal}
\bvolume{222},
\bfpage{850}
(\byear{1978})
\end{barticle}
\endbibitem

\bibitem[\protect\citeauthoryear{Horton and Hasegawa}{1994}]{Horton_Hasegawa_1994}
\begin{barticle}
\bauthor{\bsnm{Horton}, \binits{W.}},
\bauthor{\bsnm{Hasegawa}, \binits{A.}}:
\bjtitle{Chaos}
\bvolume{4},
\bfpage{227}
(\byear{1994})
\end{barticle}
\endbibitem

\bibitem[\protect\citeauthoryear{Kelvin}{1887}]{Kelvin_1887}
\begin{barticle}
\bauthor{\bsnm{Kelvin}, \binits{L.}}:
\bjtitle{Philosophical Magazine}
\bvolume{24}(\bissue{5}),
\bfpage{188}
(\byear{1887})
\end{barticle}
\endbibitem

\bibitem[\protect\citeauthoryear{Lagnado et~al.}{1984}]{Lagnado_1984}
\begin{botherref}
\oauthor{\bsnm{Lagnado}, \binits{R.}},
\oauthor{\bsnm{Phan-Thien}, \binits{N.}},
\oauthor{\bsnm{Leal}, \binits{L.}}:
Physics of Fluids
\textbf{27}
(1984)
\end{botherref}
\endbibitem

\bibitem[\protect\citeauthoryear{Lominadze et~al.}{1988}]{Lominadze_1998}
\begin{barticle}
\bauthor{\bsnm{Lominadze}, \binits{J.G.}},
\bauthor{\bsnm{Chagelishvili}, \binits{G.D.}},
\bauthor{\bsnm{Chanishvili}, \binits{R.G.}}:
\bjtitle{Pisma v Astronomicheskii Zhurnal}
\bvolume{14},
\bfpage{856}
(\byear{1988})
\end{barticle}
\endbibitem

\bibitem[\protect\citeauthoryear{Lorfing et~al.}{2023}]{Lorfing_2023}
\begin{barticle}
\bauthor{\bsnm{Lorfing}, \binits{C.Y.}},
\bauthor{\bsnm{Reid}, \binits{H.A.S.}},
\bauthor{\bsnm{Gómez-Herrero}, \binits{R.}},
\bauthor{\bsnm{Maksimovic}, \binits{M.}},
\bauthor{\bsnm{Nicolaou}, \binits{G.}},
\bauthor{\bsnm{Owen}, \binits{C.J.}},
\bauthor{\bsnm{Rodriguez-Pacheco}, \binits{J.}},
\bauthor{\bsnm{Ryan}, \binits{D.F.}},
\bauthor{\bsnm{Trotta}, \binits{D.}},
\bauthor{\bsnm{Verscharen}, \binits{D.}}:
\bjtitle{The Astrophysical Journal}
\bvolume{959}(\bissue{2}),
\bfpage{128}
(\byear{2023})
\end{barticle}
\endbibitem

\bibitem[\protect\citeauthoryear{Lovelace et~al.}{1991}]{Lovelace_1991}
\begin{barticle}
\bauthor{\bsnm{Lovelace}, \binits{R.V.E.}},
\bauthor{\bsnm{Berk}, \binits{H.L.}},
\bauthor{\bsnm{Contopoulos}, \binits{J.}}:
\bjtitle{The Astrophysical Journal}
\bvolume{379},
\bfpage{696}
(\byear{1991})
\end{barticle}
\endbibitem

\bibitem[\protect\citeauthoryear{Lovelace et~al.}{1987}]{Lovelace_1987}
\begin{barticle}
\bauthor{\bsnm{Lovelace}, \binits{R.V.E.}},
\bauthor{\bsnm{Wang}, \binits{J.C.L.}},
\bauthor{\bsnm{Sulkanen}, \binits{M.E.}}:
\bjtitle{The Astrophysical Journal}
\bvolume{315},
\bfpage{504}
(\byear{1987})
\end{barticle}
\endbibitem

\bibitem[\protect\citeauthoryear{Mahajan and Rogava}{1999}]{Mahajan_Rogava_1999}
\begin{barticle}
\bauthor{\bsnm{Mahajan}, \binits{S.M.}},
\bauthor{\bsnm{Rogava}, \binits{A.D.}}:
\bjtitle{The Astrophysical Journal}
\bvolume{518}(\bissue{2}),
\bfpage{814}
(\byear{1999})
\end{barticle}
\endbibitem

\bibitem[\protect\citeauthoryear{Mahajan et~al.}{1997}]{Mahajan_1997}
\begin{barticle}
\bauthor{\bsnm{Mahajan}, \binits{S.M.}},
\bauthor{\bsnm{Machabeli}, \binits{G.Z.}},
\bauthor{\bsnm{Rogava}, \binits{A.D.}}:
\bjtitle{The Astrophysical Journal}
\bvolume{479}(\bissue{2}),
\bfpage{129}
(\byear{1997})
\end{barticle}
\endbibitem

\bibitem[\protect\citeauthoryear{Ocker et~al.}{2021}]{Ocket_2021}
\begin{barticle}
\bauthor{\bsnm{Ocker}, \binits{S.K.}},
\bauthor{\bsnm{Cordes}, \binits{J.M.}},
\bauthor{\bsnm{Chatterjee}, \binits{S.}},
\bauthor{\bsnm{Gurnett}, \binits{D.A.}},
\bauthor{\bsnm{Kurth}, \binits{W.S.}},
\bauthor{\bsnm{Spangler}, \binits{S.R.}}:
\bjtitle{Nature Astronomy}
\bvolume{5},
\bfpage{761}
(\byear{2021})
\end{barticle}
\endbibitem

\bibitem[\protect\citeauthoryear{Osmanov et~al.}{2012}]{Osmanov_2012}
\begin{botherref}
\oauthor{\bsnm{Osmanov}, \binits{Z.}},
\oauthor{\bsnm{Rogava}, \binits{A.}},
\oauthor{\bsnm{Poedts}, \binits{S.}}:
Astronomy and Astrophysics
\textbf{19}
(2012)
\end{botherref}
\endbibitem

\bibitem[\protect\citeauthoryear{Osmanov et~al.}{2015}]{Osmanov_2015}
\begin{botherref}
\oauthor{\bsnm{Osmanov}, \binits{Z.}},
\oauthor{\bsnm{Rogava}, \binits{A.}},
\oauthor{\bsnm{Poedts}, \binits{S.}}:
New Journal of Physics
\textbf{17}
(2015)
\end{botherref}
\endbibitem

\bibitem[\protect\citeauthoryear{Parker}{1974}]{Parker_1974}
\begin{barticle}
\bauthor{\bsnm{Parker}, \binits{E.N.}}:
\bjtitle{Astronomy and Astrophysics}
\bvolume{190},
\bfpage{429}
(\byear{1974})
\end{barticle}
\endbibitem

\bibitem[\protect\citeauthoryear{Petviashvili}{1980}]{Petviashvili_1980}
\begin{barticle}
\bauthor{\bsnm{Petviashvili}, \binits{V.I.}}:
\bjtitle{Soviet Journal of Experimental and Theoretical Physics Letters}
\bvolume{32},
\bfpage{619}
(\byear{1980})
\end{barticle}
\endbibitem

\bibitem[\protect\citeauthoryear{Pike and Mason}{1998}]{Pike_Mason_1998}
\begin{barticle}
\bauthor{\bsnm{Pike}, \binits{C.D.}},
\bauthor{\bsnm{Mason}, \binits{H.E.}}:
\bjtitle{Solar Physics}
\bvolume{182}(\bissue{2}),
\bfpage{333}
(\byear{1998})
\end{barticle}
\endbibitem

\bibitem[\protect\citeauthoryear{Pneuman and Orral}{1980}]{pneuman_Orral_1980}
\begin{botherref}
\oauthor{\bsnm{Pneuman}, \binits{G.}},
\oauthor{\bsnm{Orral}, \binits{F.}}:
Physics of the Sun.
D. Reidel Publ. Co., Dordrecht, Holland
(1980)
\end{botherref}
\endbibitem

\bibitem[\protect\citeauthoryear{Poedts and Rogava}{2002}]{Poedts_Rogava_2002}
\begin{barticle}
\bauthor{\bsnm{Poedts}, \binits{S.}},
\bauthor{\bsnm{Rogava}, \binits{A.D.}}:
\bjtitle{The Astrophysical Journal}
\bvolume{385},
\bfpage{32}
(\byear{2002})
\end{barticle}
\endbibitem

\bibitem[\protect\citeauthoryear{Poedts et~al.}{1998}]{Poedts_1998}
\begin{barticle}
\bauthor{\bsnm{Poedts}, \binits{S.}},
\bauthor{\bsnm{Rogava}, \binits{A.D.}},
\bauthor{\bsnm{Mahajan}, \binits{S.M.}}:
\bjtitle{The Astrophysical Journal}
\bvolume{505}(\bissue{1}),
\bfpage{369}
(\byear{1998})
\end{barticle}
\endbibitem

\bibitem[\protect\citeauthoryear{Reid and Kontar}{2017}]{Reid_Kontar_2017}
\begin{botherref}
\oauthor{\bsnm{Reid}, \binits{H.}},
\oauthor{\bsnm{Kontar}, \binits{E.}}:
Astronomy \& Astrophysics
\textbf{598}
(2017)
\end{botherref}
\endbibitem

\bibitem[\protect\citeauthoryear{Rogava et~al.}{2003b}]{Rogava_2003b}
\begin{barticle}
\bauthor{\bsnm{Rogava}, \binits{A.D.}},
\bauthor{\bsnm{Bodo}, \binits{G.}},
\bauthor{\bsnm{Massaglia}, \binits{S.}},
\bauthor{\bsnm{Osmanov}, \binits{Z.}}:
\bjtitle{Astronomy \& Astrophysics}
\bvolume{408}(\bissue{2}),
\bfpage{401}
(\byear{2003}b)
\end{barticle}
\endbibitem

\bibitem[\protect\citeauthoryear{Rogava et~al.}{2000}]{Rogava_2000}
\begin{barticle}
\bauthor{\bsnm{Rogava}, \binits{A.D.}},
\bauthor{\bsnm{Poedts}, \binits{S.}},
\bauthor{\bsnm{Mahajan}, \binits{S.M.}}:
\bjtitle{Astronomy \& Astrophysics}
\bvolume{354},
\bfpage{749}
(\byear{2000})
\end{barticle}
\endbibitem

\bibitem[\protect\citeauthoryear{Rogava et~al.}{2001}]{Rogava_2001}
\begin{barticle}
\bauthor{\bsnm{Rogava}, \binits{A.D.}},
\bauthor{\bsnm{Poedts}, \binits{S.}},
\bauthor{\bsnm{Mahajan}, \binits{S.M.}}:
\bjtitle{Journal of Computational Acoustics}
\bvolume{09}(\bissue{03}),
\bfpage{869}
(\byear{2001})
\end{barticle}
\endbibitem

\bibitem[\protect\citeauthoryear{Rogava et~al.}{2003a}]{Rogava_2003a}
\begin{barticle}
\bauthor{\bsnm{Rogava}, \binits{A.D.}},
\bauthor{\bsnm{Mahajan}, \binits{S.M.}},
\bauthor{\bsnm{Bodo}, \binits{G.}},
\bauthor{\bsnm{S.}, \binits{M.}}:
\bjtitle{Astronomy \& Astrophysics}
\bvolume{399},
\bfpage{421}
(\byear{2003}a)
\end{barticle}
\endbibitem

\bibitem[\protect\citeauthoryear{Rogava et~al.}{2007}]{Rogava_2007}
\begin{barticle}
\bauthor{\bsnm{Rogava}, \binits{A.}},
\bauthor{\bsnm{Gogoberidze}, \binits{G.}},
\bauthor{\bsnm{Poedts}, \binits{S.}}:
\bjtitle{The Astrophysical Journal}
\bvolume{664}(\bissue{2}),
\bfpage{1221}
(\byear{2007})
\end{barticle}
\endbibitem

\bibitem[\protect\citeauthoryear{Rogava and Khujadze}{1997}]{Rogava_Khujadze_1997}
\begin{barticle}
\bauthor{\bsnm{Rogava}, \binits{A.D.}},
\bauthor{\bsnm{Khujadze}, \binits{G.R.}}:
\bjtitle{General Relativity and Gravitation}
\bvolume{29},
\bfpage{345}
(\byear{1997})
\end{barticle}
\endbibitem

\bibitem[\protect\citeauthoryear{Rogava et~al.}{1998}]{Rogava_1998}
\begin{barticle}
\bauthor{\bsnm{Rogava}, \binits{A.D.}},
\bauthor{\bsnm{Chagelishvili}, \binits{G.D.}},
\bauthor{\bsnm{Mahajan}, \binits{S.M.}}:
\bjtitle{Physical Reiew E}
\bvolume{57},
\bfpage{7103}
(\byear{1998})
\end{barticle}
\endbibitem

\bibitem[\protect\citeauthoryear{Rogava et~al.}{1999}]{Rogava_1999}
\begin{barticle}
\bauthor{\bsnm{Rogava}, \binits{A.D.}},
\bauthor{\bsnm{Poedts}, \binits{S.}},
\bauthor{\bsnm{Heirman}, \binits{S.}}:
\bjtitle{Monthly Notices of the Royal Astronomical Society}
\bvolume{307}(\bissue{4}),
\bfpage{31}
(\byear{1999})
\end{barticle}
\endbibitem

\bibitem[\protect\citeauthoryear{Saffman}{1993}]{Saffman_1993}
\begin{bbook}
\bauthor{\bsnm{Saffman}, \binits{P.G.}}:
\bbtitle{Vortex Dynamics}.
\bsertitle{Cambridge Monographs on Mechanics}.
\bpublisher{Cambridge University Press}, \blocation{???}
(\byear{1993})
\end{bbook}
\endbibitem

\bibitem[\protect\citeauthoryear{Tagger et~al.}{1992}]{Tagger_1992}
\begin{barticle}
\bauthor{\bsnm{Tagger}, \binits{M.}},
\bauthor{\bsnm{Pellat}, \binits{R.}},
\bauthor{\bsnm{Coroniti}, \binits{F.V.}}:
\bjtitle{Astrophysical Journal}
\bvolume{393},
\bfpage{708}
(\byear{1992})
\end{barticle}
\endbibitem

\bibitem[\protect\citeauthoryear{Toomre}{1969}]{Toomre_1969}
\begin{barticle}
\bauthor{\bsnm{Toomre}, \binits{A.}}:
\bjtitle{Astrophysical Journal}
\bvolume{158},
\bfpage{899}
(\byear{1969})
\end{barticle}
\endbibitem

\bibitem[\protect\citeauthoryear{Trefethen et~al.}{1993}]{Trefethen_1993}
\begin{barticle}
\bauthor{\bsnm{Trefethen}, \binits{L.N.}},
\bauthor{\bsnm{Trefethen}, \binits{A.E.}},
\bauthor{\bsnm{Reddy}, \binits{S.C.}},
\bauthor{\bsnm{Driscoll}, \binits{T.A.}}:
\bjtitle{Science}
\bvolume{261}(\bissue{5121}),
\bfpage{578}
(\byear{1993})
\end{barticle}
\endbibitem

\end{thebibliography}

\end{document}